\documentclass[]{tJOT2e}
\begin{document}

\markboth{M. Gonzalez}{Journal of Turbulence}

\articletype{RESEARCH ARTICLE - Revised version}

\title{Analysis of scalar dissipation
in terms of \\ vorticity geometry
in isotropic turbulence}

\author{M. Gonzalez$^{\ast}$\thanks{$^\ast$ Email: Michel.Gonzalez@coria.fr
\vspace{6pt}} \\\vspace{6pt}  {\em{CNRS, UMR 6614/CORIA}}
\\ {\em{Site universitaire du Madrillet}}
\\ {\em{76801 Saint-Etienne du Rouvray, France}}
\\\vspace{6pt}\received{v3.2 released February 2009} }

\maketitle

\begin{abstract}

The mechanisms promoting scalar dissipation through scalar
gradient production are scrutinized in terms of vorticity
alignment with respect to strain principal axes.
For that purpose, a stochastic Lagrangian model for
the velocity gradient tensor and the scalar gradient vector
is used. 
The model results show that the major part of scalar
dissipation occurs for stretched vorticity, namely
when the vorticity vector
aligns with the extensional
and intermediate strain eigenvectors. More specifically,
it appears that
the mean scalar dissipation
is well represented by the sample defined by alignment with
the extensional strain,
while
the most intense scalar dissipation 
is promoted by the set of events for which vorticity aligns
with the intermediate strain.
This difference is explained by 
rather subtle mechanisms involving the
statistics of both the strain intensities
and the scalar gradient alignment 
resulting from these special alignments of vorticity.
The analysis allowing for the local flow structure 
confirms
the
latter scenario for both the strain-
and rotation-dominated events.
However, 
despite the prevailing role of strain in
promoting scalar dissipation,
the difference in the level of scalar dissipation
when vorticity aligns with either the extensional or
the intermediate strain mostly arises from rotation-dominated
events.

\begin{keywords}
mixing; scalar dissipation; scalar gradient;
vorticity geometry

\end{keywords}

\end{abstract}

\section{Introduction}
\label{sec1}

The finest level at which micromixing in fluid flows
can be investigated is determined by the local gradient of
a scalar.
The gradient indeed rules molecular diffusion, but also gives
a precise insight into the small-scale structure of scalar
fields and mixing patterns.
Actually, it is the mean dissipation rate of the energy of
scalar fluctuations,
$ D \langle \bm{G}^2 \rangle $,          
the so-called scalar dissipation
-- with $ D $ the molecular diffusivity and $ \bm{G} $
the scalar gradient --, which reveals the efficiency of micromixing.
Modelling small-scale mixing
in process and chemical engineering \cite{BP95}
or
in combustion flow computation \cite{B04}
thus needs to understand the very mechanisms
of scalar gradient production.
On the basic level, the study of the scalar gradient is
relevant
to the general problem of vector transport in fluid
flows \cite{S63} including the kinematics of vectors defining
material lines or surfaces \cite{GP90},
the vorticity vector properties \cite{NP98,Lal05}, the
dynamics of the vorticity gradient in two-dimensional flows 
\cite{DB03}
as well
as the production of the magnetic field by the motion of a conducting 
fluid
\cite{FB12}.

A number of studies have tackled the connection 
of
the detailed features of the scalar gradient and of the scalar 
dissipation with the properties of the flow field
determined by the local velocity gradients
\cite{K85,Aal87,RM92,P94,HS94,BD96,BD98,W00,Val01,Bal03,Gal07,
Aal09}.
Clearly, in scalar gradient production,
strain
-- through its
intensity, persistence and the respective alignments
of scalar gradient and principal axes --
is the chief mechanism,
while vorticity, at least its magnitude, is immaterial. 
Because of the tight interaction between 
strain and vorticity \cite{NP98,T00}, however,
vorticity properties are likely to be indirectly involved.
In fact, the role of strain in the vicinity of vorticity
structures has early been established \cite{K85,Aal87,RM92}.
Models based on stretched vortices \cite{PL01} have also been
shown to reproduce the physics of scalar transport and mixing
in turbulent flows.
More recently the influence of vortical structures on mixing
has been clearly shown \cite{Kal11}.

The present 
work, too,
has to do with
the relationship between mixing properties
and the local features of the flow field.
The production of scalar dissipation,
and thus the mechanisms of micromixing, are probed through the kinematic
features of the scalar gradient in terms of vorticity geometry.
Since
vorticity alignments
arise from 
the dynamics of the velocity field
and
are closely connected to
the inner, detailed structure of turbulent flows
\cite{NP98,Lal05,T98},
alignment of vorticity with respect to strain principal axes
is especially considered. 

This article 
reports an extension of the findings
presented in reference \cite{GP11}. 
The stochastic Lagrangian model used in the study is
described in Section \ref{sec2} and its ability to predict
statistics conditioned on vorticity alignments is
checked
in Section \ref{sec3}. Making use of the model results, the
scrutiny
of scalar dissipation in terms of vorticity geometry,
including the
analysis based on local flow structure,
is achieved in Section \ref{sec4}.
Conclusion
is drawn in Section \ref{sec5}.

\section{Stochastic Lagrangian model for the velocity and scalar gradients}
\label{sec2}

\subsection{Modelled equations}
\label{sec2.1}

The model for the velocity gradient tensor has been derived by
Chevillard and Meneveau \cite{CM06}
and has been shown
to predict the essential geometric properties and anomalous 
scalings of incompressible, isotropic turbulence
\cite{CM06,Cal08}.
Starting from an Eulerian-Lagrangian change of variables and
using the Recent Fluid Deformation Approximation the modelled equation
for the velocity gradient tensor, $ {\bm A} $, is derived as
\begin{equation}
\label{eq1}
d{\bm A}
=
\left(
-{\bm A}^2
+
\frac{Tr({\bm A}^2)}{Tr({\bm C}_{\tau_{\eta}}^{-1})}
 {\bm C}_{\tau_{\eta}}^{-1}
-
\frac{Tr({\bm C}_{\tau_{\eta}}^{-1})}{3 T} {\bm A}
\right) dt
+
{\left(\frac{2}{T} \right)}^{1/2} d{\bm W}
\end{equation}
in which 
$ T $ is the integral time scale and
$ {\bm C}_{\tau_{\eta}} $
is a model for the Cauchy-Green tensor,
$ {\bm C}_{\tau_{\eta}}
=
\exp({\tau_{\eta} {\bm A}})
\exp({\tau_{\eta} {\bm A}^T}) $,
where
$ \tau_{\eta} $ is the Kolmogorov time scale.
Forcing is ensured by the
increment of a tensorial Wiener process,
$ d{\bm W} = dt^{1/2} {\bm \zeta} $,
where $ {\bm \zeta} $ is a tensorial, Gaussian delta-correlated noise
with
$ \langle \zeta_{ij} \rangle = 0 $ and
$ \langle \zeta_{ij} \zeta_{kl} \rangle =
2 \delta_{ik} \delta_{jl} - 1/2 \delta_{ij} \delta_{kl}
-1/2 \delta_{il} \delta_{jk} $.

This model has been extended to the gradient of a passive scalar
\cite{G09}.
The modelled equation for the scalar gradient is
written
\begin{equation}
\label{eq2}
d {\bm G}
=
- \left(
{\bm A}^T {\bm G}
+
\frac{Tr({\bm C}_{\tau_{\eta}}^{-1})}{3 T_{\theta}} {\bm G}
\right) dt
+
{\left(\frac{2}{T_{\theta}} \right)}^{1/2} d{\bm W}_{\bm G}
\end{equation}
where $ T_{\theta} $ is the scalar integral time scale and
$ d{\bm W}_{\bm G} = dt^{1/2} {\bm \xi} $ is 
the increment of a Wiener process
where $ {\bm \xi} $ is a vectorial, Gaussian noise such that
$ \langle \xi_i \rangle = 0 $ and
$ \langle \xi_i \xi_j \rangle = \delta_{ij} $.

In the model represented by Eqs. (\ref{eq1}) and (\ref{eq2})
stretching is exactly taken into account,
while
models are devised for
the pressure Hessian -- second term of Eq. (\ref{eq1}) --,
viscous effects -- third term of Eq. (\ref{eq1}) --
and molecular diffusion
-- second term of
Eq. (\ref{eq2}).
Meneveau has given a detailed discussion
on this class of stochastic Lagrangian models
\cite{M11}.

\subsection{Numerical solution}
\label{sec2.2}

Time scales are normalised by the integral time scale
($ T = 1 $). As in reference \cite{G09}
the Kolmogorov time scale
and the scalar integral time scale
are respectively prescribed
as
$ \tau_{\eta} = 0.1 $
-- which corresponds to a Taylor microscale Reynolds number,
$ {\rm Re}_{\lambda} $,
close
to 150 \cite{Cal08} -- 
and $ T_{\theta} = 0.4 $.

Equations (\ref{eq1}) and (\ref{eq2})
are solved using a second-order predictor-corrector scheme
\cite{WP97}.
The calculation is run for $ 2 \times 10^5 T $ with time
step $ 10^{-2} $ and the statistics of the
velocity and scalar gradients 
are derived from their respective stationary time signals.

\section{Predictions of statistical features conditioned on
\label{sec3}
vorticity properties}

\begin{figure}[!h]
\begin{center}
\resizebox*{12cm}{!}{\includegraphics{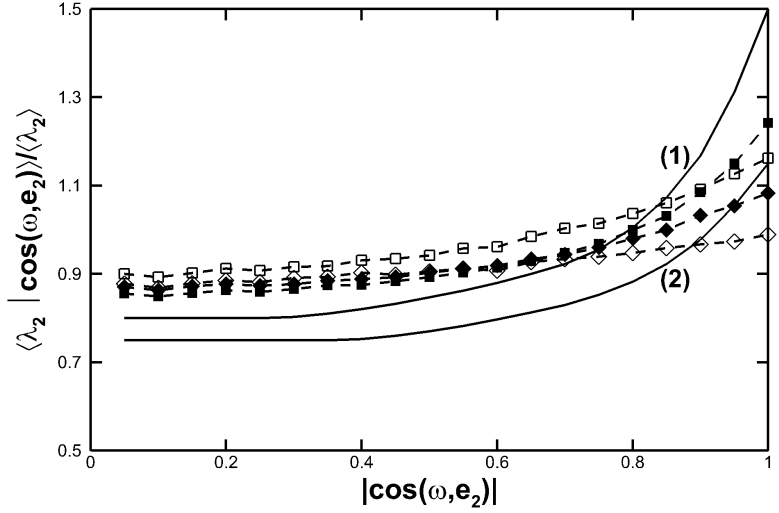}}\\
\caption{
Normalised mean intermediate strain eigenvalue conditioned on the alignment
of vorticity with the intermediate strain eigenvector; solid lines:
DNS of Tsinober {\em et al.} \cite{Tal97}; symbols: model results;
(1) and squares: whole field; (2) and diamonds: 
$ \bm{\omega}^2 < \langle \bm{\omega}^2 \rangle $;
full symbols: $ {\rm Re}_{\lambda} \simeq 150 $; 
open symbols: $ {\rm Re}_{\lambda} \simeq 75 $.} 
\label{fig1}
\end{center}
\end{figure}

\begin{figure}[!h]
\begin{center}
\resizebox*{12cm}{!}{\includegraphics{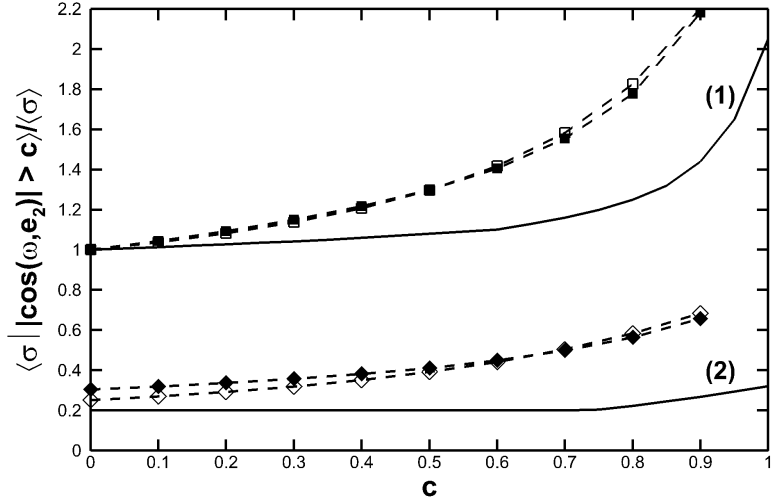}}\\
\caption{
Mean production rate of enstrophy conditioned on strong alignment
of vorticity with the intermediate strain eigenvector; 
$ \sigma = \bm{\omega}^2 [\lambda_1 \cos^2(\bm{\omega},\bm{e}_1)
+
\lambda_2 \cos^2(\bm{\omega},\bm{e}_2)
+
\lambda_3 \cos^2(\bm{\omega},\bm{e}_3)] $;
solid lines:
DNS of Tsinober {\em et al.} \cite{Tal97}; symbols: model results;
(1) and squares: whole field; (2) and diamonds: 
$ \bm{\omega}^2 < \langle \bm{\omega}^2 \rangle $;
full symbols: $ {\rm Re}_{\lambda} \simeq 150 $; 
open symbols: $ {\rm Re}_{\lambda} \simeq 75 $.} 
\label{fig2}
\end{center}
\end{figure}

\begin{figure}[!h]
\begin{center}
\resizebox*{12cm}{!}{\includegraphics{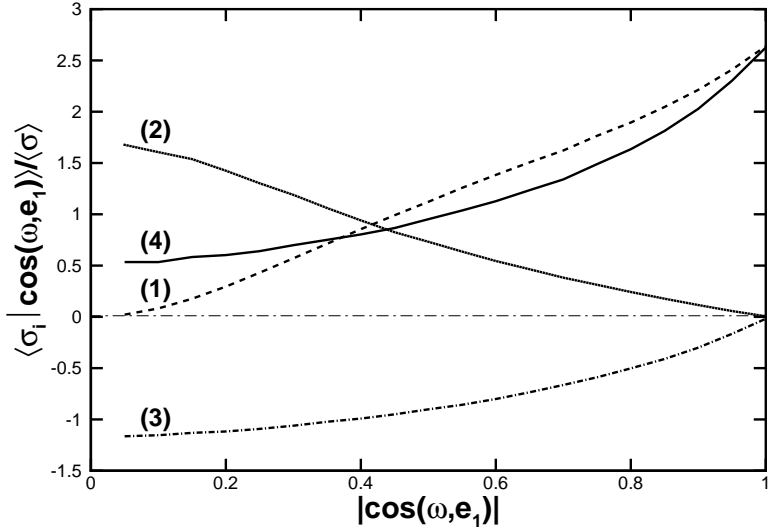}}\\
\caption{
Normalised mean enstrophy production terms conditioned on the alignment
of vorticity with the extensional strain eigenvector; 
(1): average of $ \sigma_1 = \bm{\omega}^2 \lambda_1 \cos^2(\bm{\omega},\bm{e}_1) $;
(2): average of $ \sigma_2 = \bm{\omega}^2 \lambda_2 \cos^2(\bm{\omega},\bm{e}_2) $;
(3): average $ \sigma_3 = \bm{\omega}^2 \lambda_3 \cos^2(\bm{\omega},\bm{e}_3) $;
(4): average of $ \sigma = \sigma_1 + \sigma_2 + \sigma_3 $;
the figure compares well with Fig. 11 of Tsinober {\em et al.}
\cite{Tal97}.}
\label{fig3}
\end{center}
\end{figure}

The model 
retrieves 
the main features of 
the scalar gradient
statistics and kinematics 
\cite{G09}, namely the non-gaussian properties of the scalar
gradient components, the probability density functions (p.d.f.'s)
of the production of scalar gradient norm, the statistical
alignments with respect to strain principal axes and vorticity
as well as more subtle features already underlined in two-dimensional
turbulence \cite{Lal99} such as the existence of special preferential
alignments. 
It has also been shown to reproduce the statistics of the
scalar gradient in rotating turbulence \cite{L11}.
This Lagrangian approach has been used to model
the evolution of the turbulent magnetic field as well
\cite{Hal11}.
Additional assessment of the model in connection with the present study
relates to statistics conditioned on vorticity alignments. The latter are
taken from the direct numerical simulations (DNS) by Tsinober {\em et al.}
\cite{Tal97} and comparisons with the model predictions
are made
in Figs. \ref{fig1} - \ref{fig3}.

The strain eigenvalues are denoted by $ \lambda_i $ and the corresponding
eigenvectors by $ \bm{e}_i $; the $ \lambda_i$'s are such that
$ \lambda_3 < \lambda_2 < \lambda_1 $ with $ \lambda_1 + \lambda_2 + \lambda_3 = 0 $
and $ \bm{e}_1 $, $ \bm{e}_2 $ and $ \bm{e}_3 $ define, respectively, the extensional,
intermediate and compressional strain principal axes.
Figure \ref{fig1} displays the normalised average of the intermediate strain eigenvalue
conditioned on the alignment of vorticity, $ \bm{\omega} $, with respect to the intermediate
strain eigenvector and shows that the increase of $ \lambda_2 $ with
$ |\cos(\bm{\omega},\bm{e}_2)| $ is reasonably predicted by the model for
the whole field as well as for small vorticity.

Figure \ref{fig2} relates to enstrophy production.
The model overpredicts the production rate of enstrophy for the strongest
alignments between vorticity and the intermediate eigenvector, but
displays the right trend for both the whole field and small vorticity,
namely the rise of the enstrophy production
rate as the alignment gets tighter.

The differences between model predictions and DNS data
shown in Figs. \ref{fig1} and \ref{fig2}
are not explained by a Reynolds number dependence.
In fact, Tsinober {\em et al.} \cite{Tal97} 
suggest that their DNS results -- although they were derived at
$ {\rm Re}_{\lambda} \simeq 85 $ -- 
are likely to be almost 
Reynolds-number independent.
The model results, displayed for 
$ {\rm Re}_{\lambda} \simeq $ 150 and 75, are consistent with
this surmise.

In agreement with the numerical
simulations of Tsinober {\em et al.} \cite{Tal97} -- see their Fig. 11 --, 
the model also
correctly predicts the normalised mean enstrophy production terms
conditioned on the alignment between vorticty and the extensional
strain eigenvector (Fig. \ref{fig3}).

\section{Analysis of scalar dissipation conditioned on vorticity
\label{sec4}
geometry}
\subsection{Production of scalar gradient
vs. vorticity alignments}
\label{sec4.1}
As scalar dissipation is in direct proportion to the square of the scalar
gradient norm,
production mechanisms of scalar gradient reveal the way in
which it is promoted by the flow field.
Using the model, 
we specifically focus on the case of significant alignment of
vorticity with the strain principal axes. The latter is defined by
$ |\cos(\bm{\omega},\bm{e}_i)| \geq c $ where $ c $ is a threshold
spanning the range 0.7 to 0.99 and alignment of vorticity with a strain
eigenvector, $ \bm{e}_i $, is denoted by $ \bm{\omega} // \bm{e}_i $.
The major part -- 92\% -- of the results corresponds to vorticity
making an angle smaller than $ 46^{\circ} $ 
($ |\cos(\bm{\omega},\bm{e}_i)| \geq 0.7 $)
with one of the strain eigenvectors; more precisely,
22\% for $ i = 1 $, 55\% for $ i= 2 $ and 15\% for $ i = 3 $.
Vorticity aligning with
$ \bm{e}_2 $ is mostly stretched;
the intermediate eigenvalue, $ \lambda_2 $, is positive in more
than $ 80\% $ of the 
$ \bm{\omega} // \bm{e}_2 $-events
-- 83\% for $ c = 0.7 $, 86\% for $ c = 0.99 $.

\begin{figure}[!h]
\begin{center}
\begin{minipage}{140mm}
\subfigure[]{
\resizebox*{7cm}{!}{\includegraphics{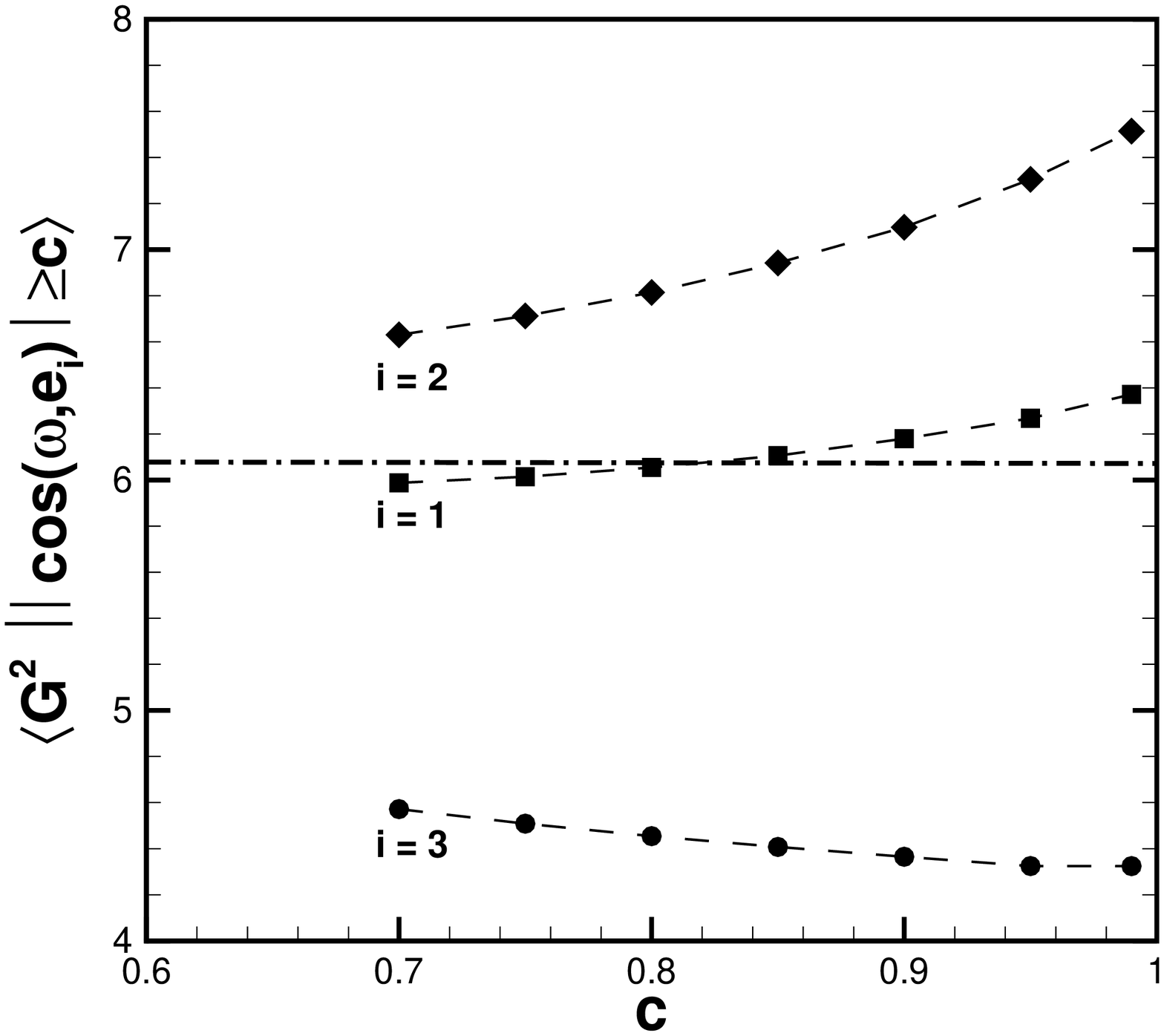}}}%
\subfigure[]{
\resizebox*{7cm}{!}{\includegraphics{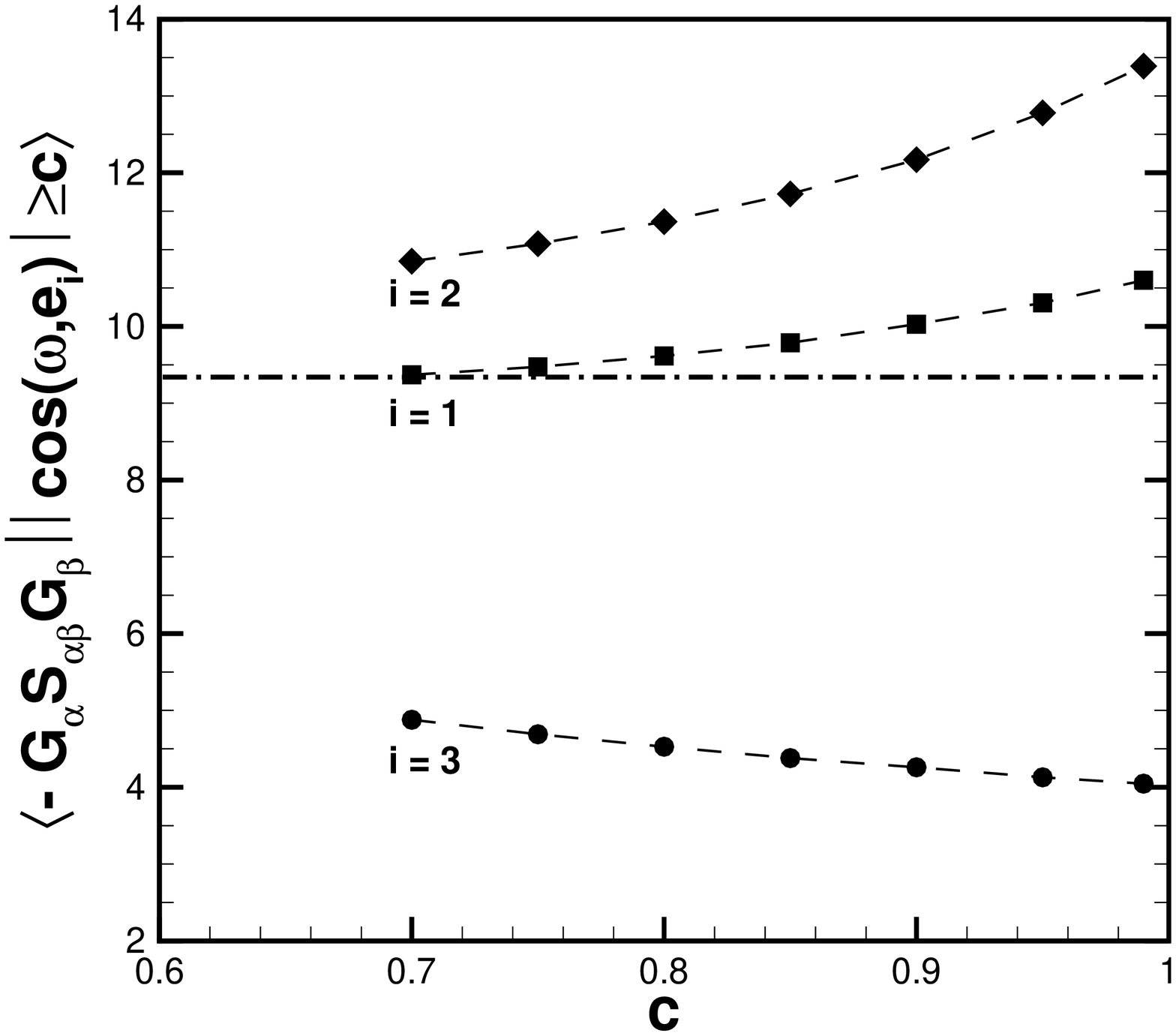}}}%
\caption{                         %
Mean square of scalar gradient norm (a)
and production term (b)
conditioned on vorticity
alignments with respect to the strain eigenvectors,
$ {\bm e}_i $;
$ i = 1 $: extensional,
$ i = 2 $: intermediate,
$ i = 3 $: compressional;
dash-dotted line: unconditioned average.}%
\label{fig4}
\end{minipage}
\end{center}
\end{figure}

Figure \ref{fig4} clearly shows the dependence of scalar gradient
production on vorticity alignments. The averages of the scalar gradient
norm, $ \bm{G}^2 $, and of its production term,
$ - G_{\alpha} S_{\alpha \beta} G_{\beta} $ (where $ \bm{S} $ is
the strain tensor),
conditioned on $ |\cos(\bm{\omega},\bm{e}_i)| \geq c $
display the same trend: their largest values correspond to strong alignment
of vorticity with $ \bm{e}_2 $, while mean values 
conditioned on alignment with $ \bm{e}_1 $
are closer to the unconditioned averaged values.
Alignment of vorticity with the compressional eigenvector,
$ \bm{e}_3 $, corresponds to the smallest production. These results thus 
suggest that the most intense scalar dissipation occurs for
$ \bm{\omega} // \bm{e}_2 $, while the mean scalar dissipation 
is rather well represented by the set of events for which
$ \bm{\omega} // \bm{e}_1 $.

\begin{figure}[!h]
\begin{center}
\begin{minipage}{140mm}
\subfigure[]{
\resizebox*{7cm}{!}{\includegraphics{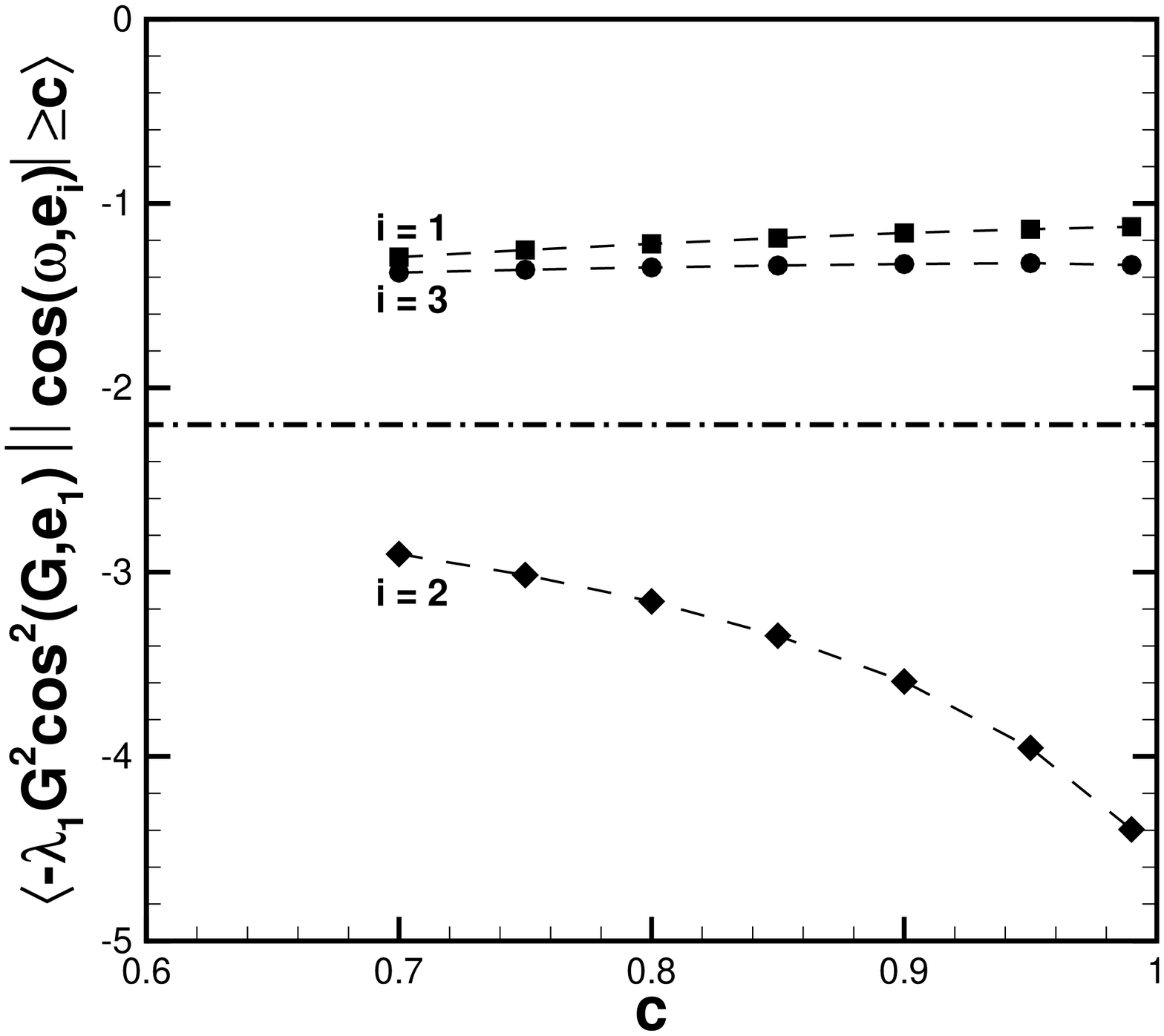}}}%
\subfigure[]{
\resizebox*{7cm}{!}{\includegraphics{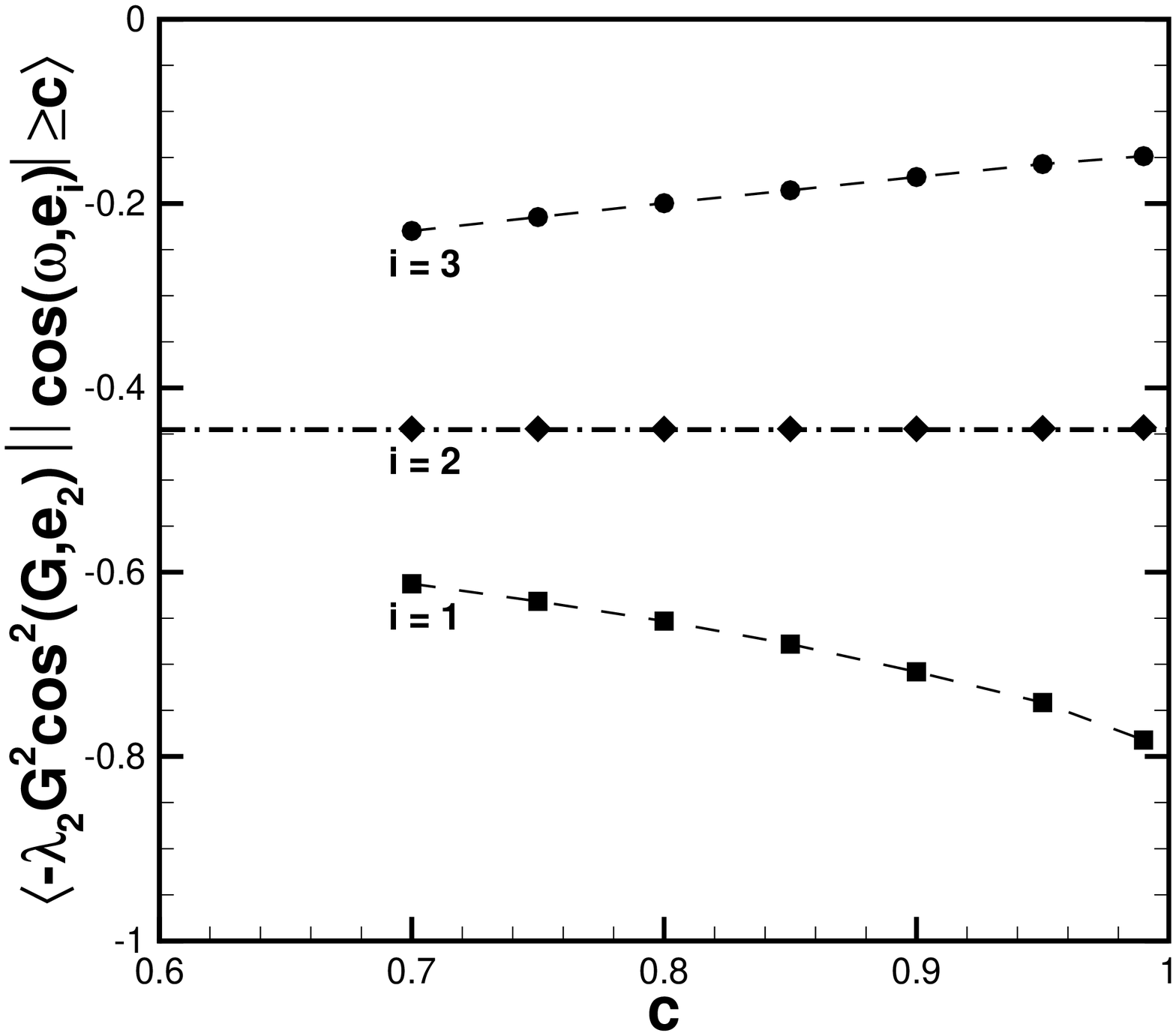}}}%
\end{minipage}
\begin{minipage}{140mm}
\begin{center}
\subfigure[]{
\resizebox*{7cm}{!}{\includegraphics{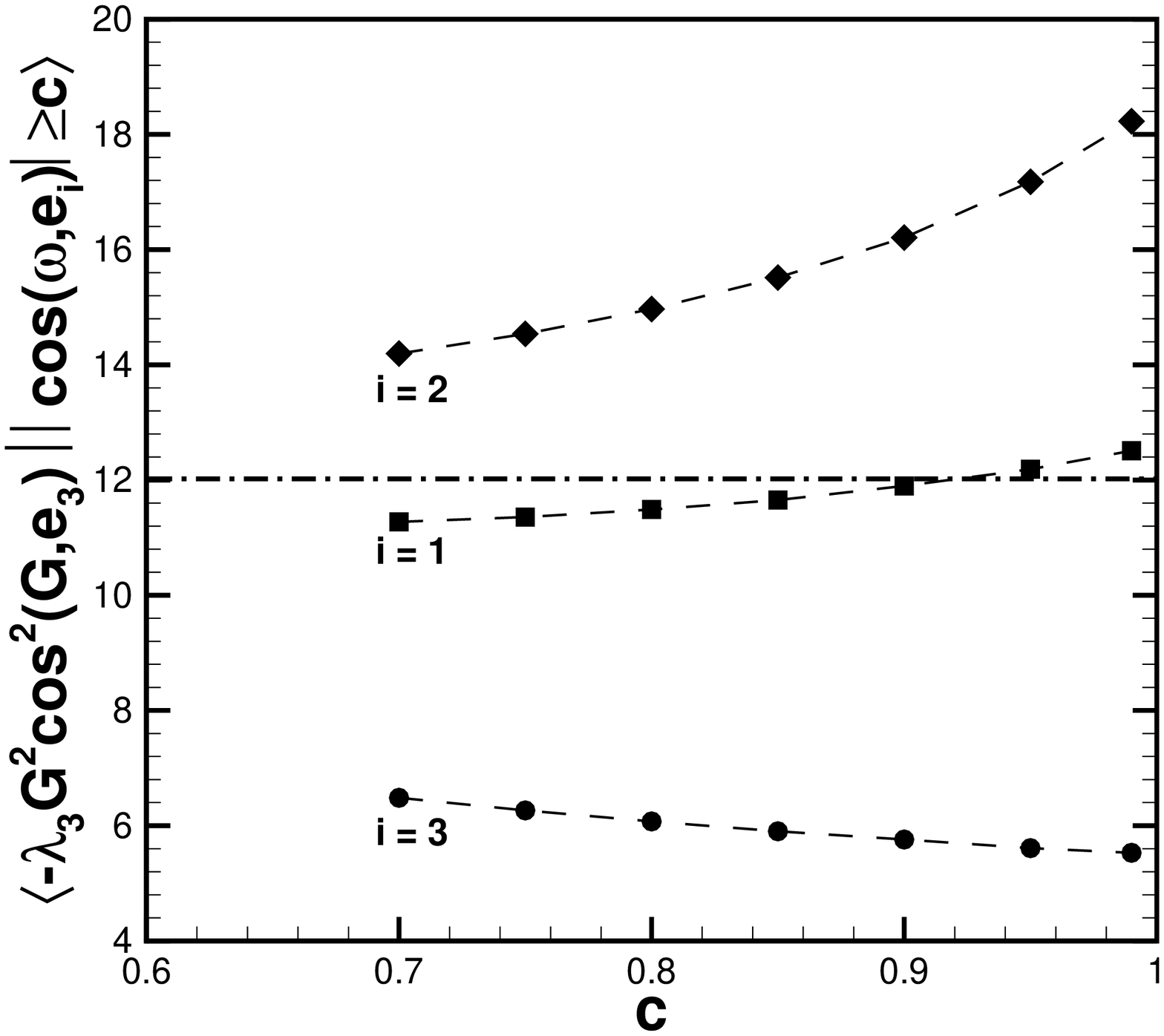}}}%
\end{center}
\caption{Individual production terms of the mean square
of scalar gradient norm conditioned on vorticity alignments
with respect to the strain eigenvectors, $ {\bm e}_i $;
(a) production by the extensional strain;
(b) by the intermediate strain;
(c) by the compressional strain;
dash-dotted line: unconditioned average.}%
\label{fig5}
\end{minipage}
\end{center}
\end{figure}

As shown in Fig. \ref{fig5},
when vorticity aligns with a strain eigenvector
the intermediate strain causes destruction
of the scalar gradient, and 
the compressional and
the extensional strains, as expected, essentially cause
production and destruction, respectively.
In addition, the differences in scalar gradient production
resulting from the vorticity alignments
stem from rather subtle mechanisms.
From Fig. \ref{fig5} it is clear that
both the weakest production and destruction 
occur for $ \bm{\omega} // \bm{e}_3 $.
Production by the compressional strain as well as destruction by the
extensional strain are the largest for $ \bm{\omega} // \bm{e}_2 $,
while in this latter case destruction by the intermediate strain coincides with its
unconditioned value. For $ \bm{\omega} // \bm{e}_1 $,
production by the compressional strain is close to the unconditioned
mean value, destruction by the extensional strain is weak and
destruction by the intermediate strain is the largest.

\begin{figure}[!h]
\begin{center}
\begin{minipage}{140mm}
\subfigure[]{
\resizebox*{7cm}{!}{\includegraphics{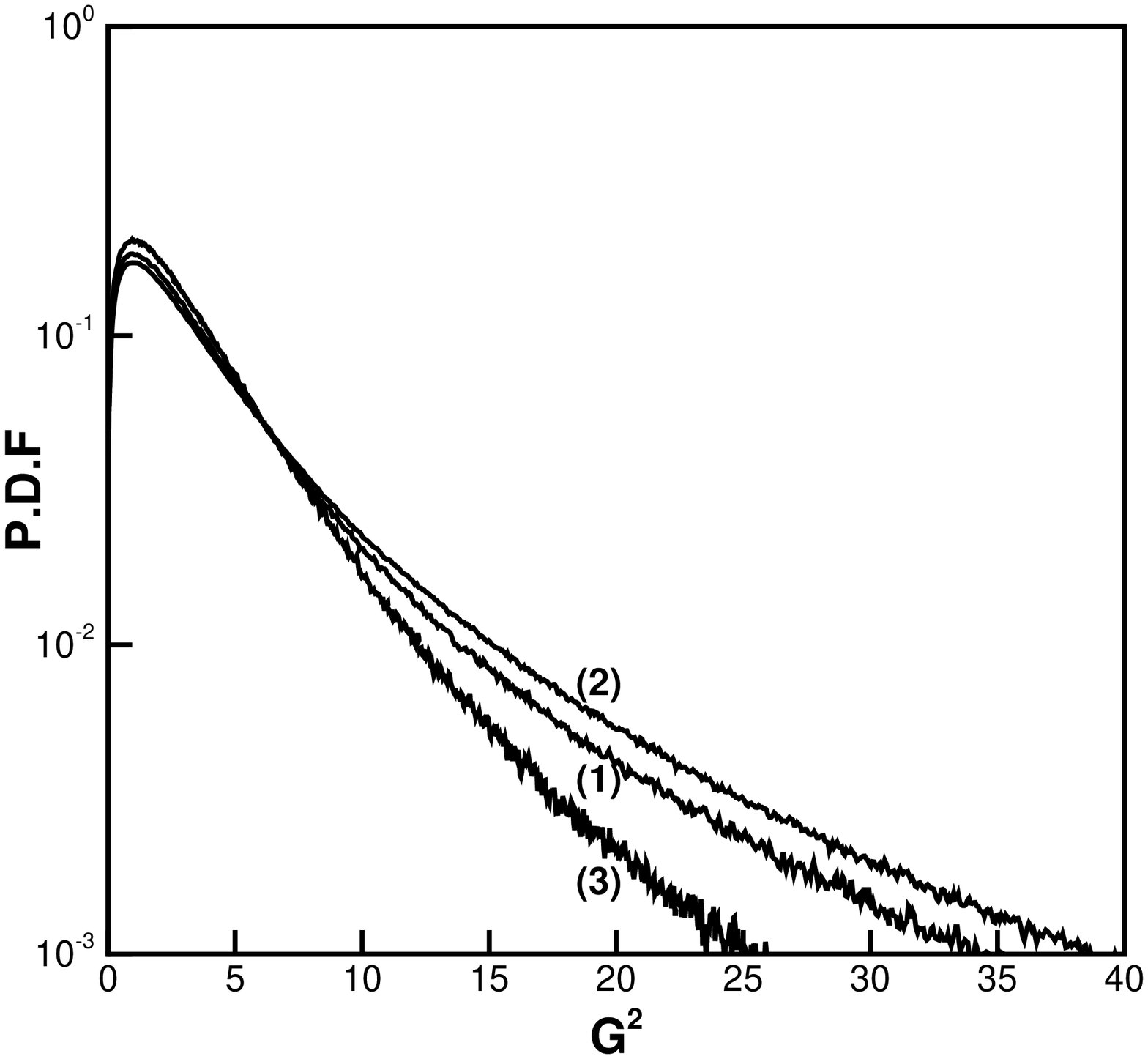}}}%
\subfigure[]{
\resizebox*{7cm}{!}{\includegraphics{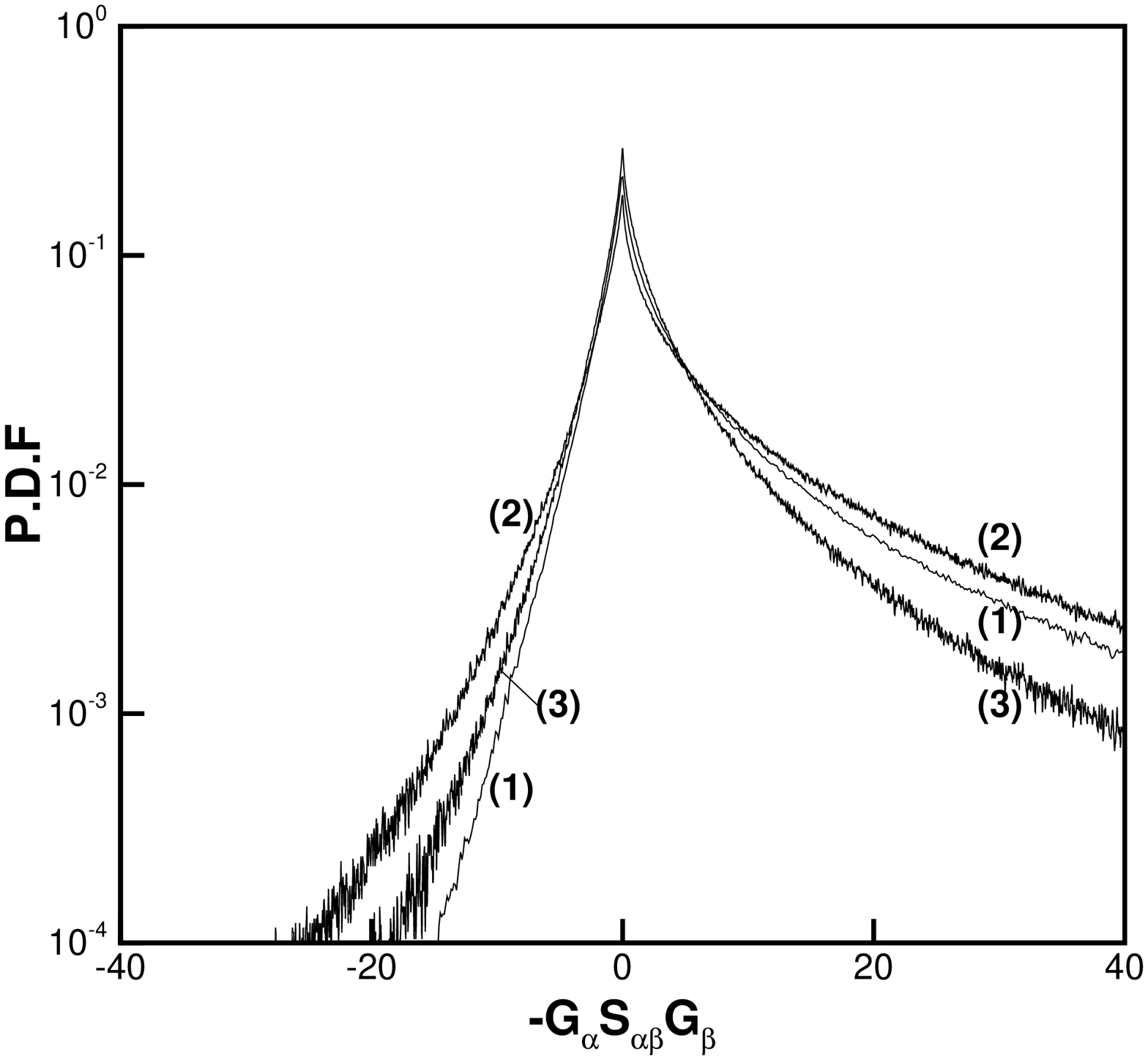}}}%
\caption{                         %
P.d.f's of (a): $ \bm{G}^2 $ 
and (b):
production term, $ -G_{\alpha}S_{\alpha \beta}G_{\beta} $, 
conditioned on vorticity alignments
defined as $ |\cos(\bm{\omega},\bm{e}_i)| \geq 0.8 $;
(1): $ \bm{\omega} // \bm{e}_1 $;
(2): $ \bm{\omega} // \bm{e}_2 $;
(3): $ \bm{\omega} // \bm{e}_3 $.}
\label{fig6}
\end{minipage}
\end{center}
\end{figure}

These results are supported by the p.d.f's of $ \bm{G}^2 $
and of the production term $ -G_{\alpha}S_{\alpha \beta}G_{\beta} $
(Fig. \ref{fig6}) conditioned on vorticity alignments. In particular, it is clear that
extreme values of $ \bm{G}^2 $ are the most probable when $ \bm{\omega} // \bm{e}_2 $.
It also appears that both the largest destruction and production of the
scalar gradient occur in this case.

\subsection{Analysis of the production mechanisms}
\begin{figure}[!h]
\begin{center}
\begin{minipage}{140mm}
\subfigure[]{
\resizebox*{7cm}{!}{\includegraphics{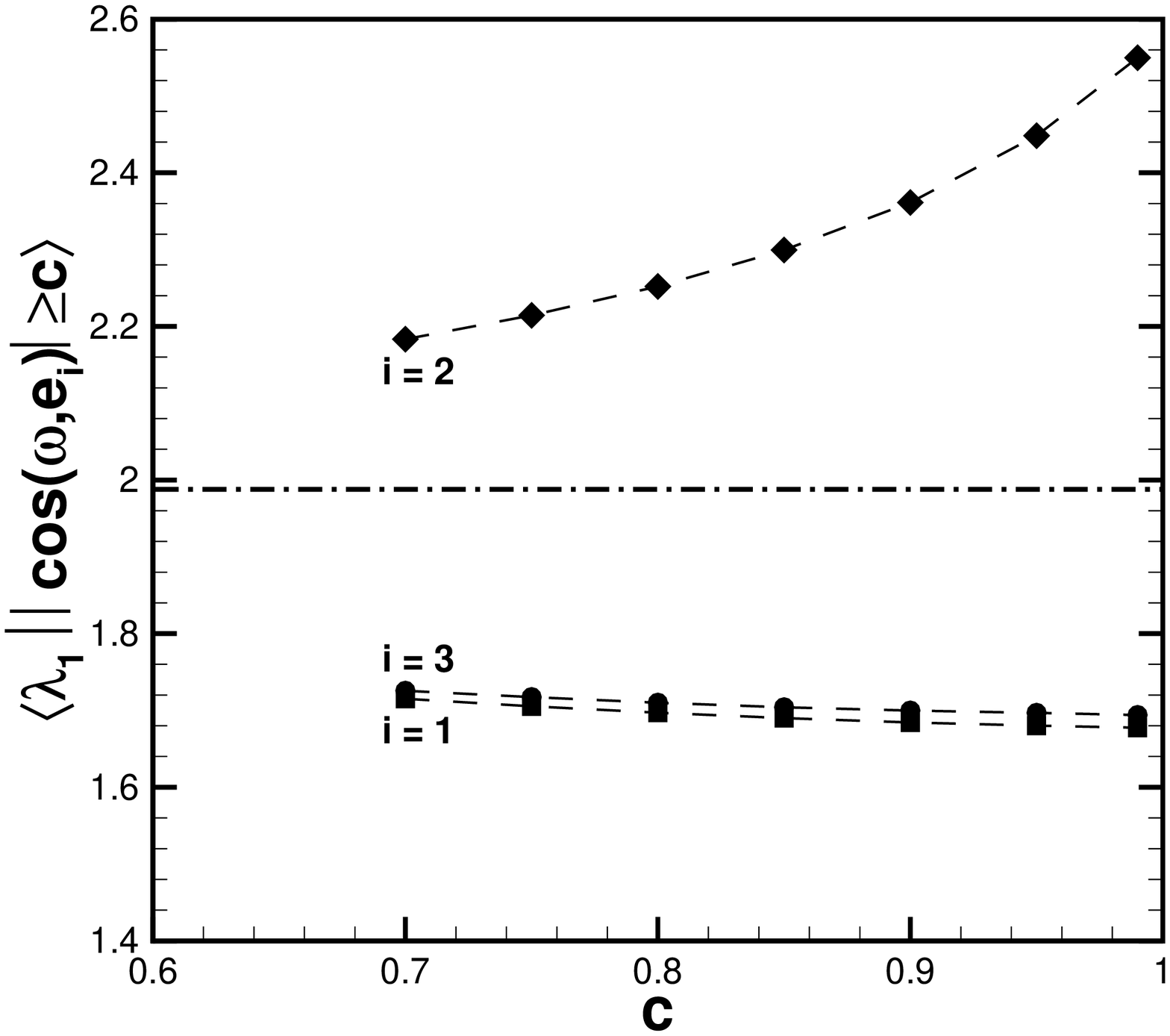}}}%
\subfigure[]{
\resizebox*{7cm}{!}{\includegraphics{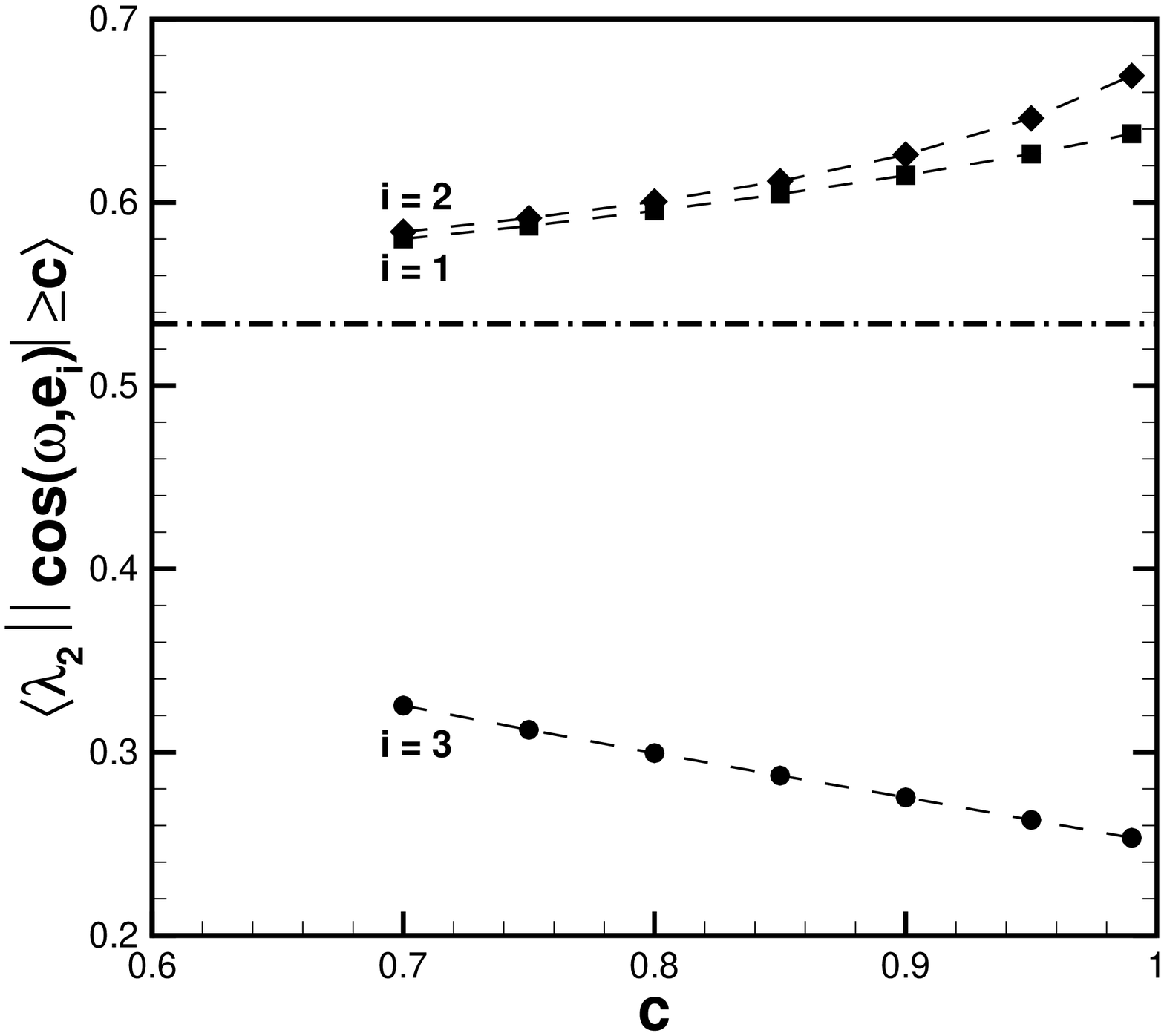}}}%
\end{minipage}
\begin{minipage}{140mm}
\begin{center}
\subfigure[]{
\resizebox*{7cm}{!}{\includegraphics{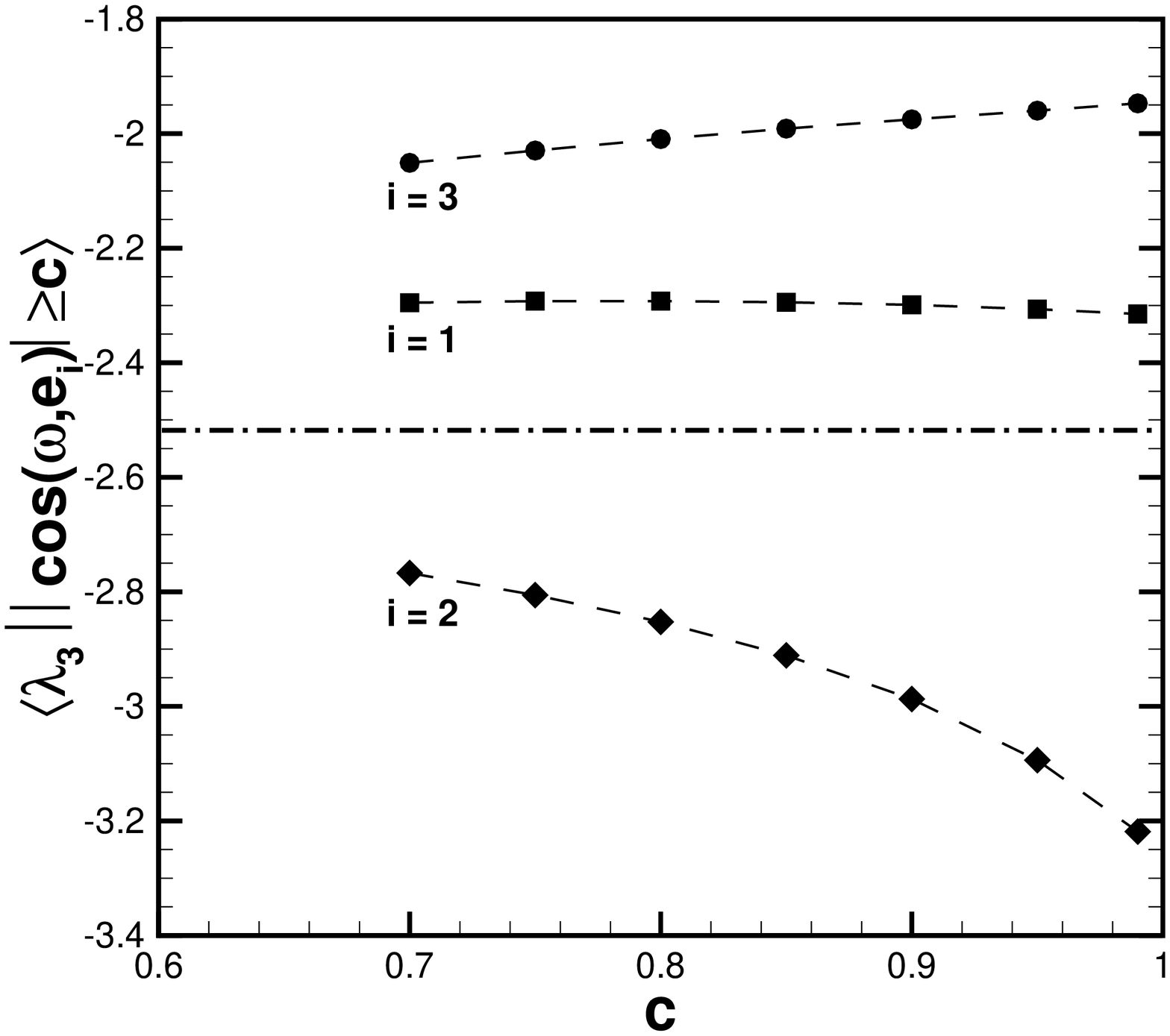}}}%
\end{center}
\caption{Mean strain eigenvalues
conditioned on vorticity alignments
with respect to the strain eigenvectors, $ {\bm e}_i $;
(a) extensional strain eigenvalue;
(b) intermediate strain eigenvalue;
(c) compressional strain eigenvalue;
dash-dotted line: unconditioned average.}%
\label{fig7}
\end{minipage}
\end{center}
\end{figure}

\begin{figure}[!h]
\begin{center}
\begin{minipage}{140mm}
\subfigure[]{
\resizebox*{7cm}{!}{\includegraphics{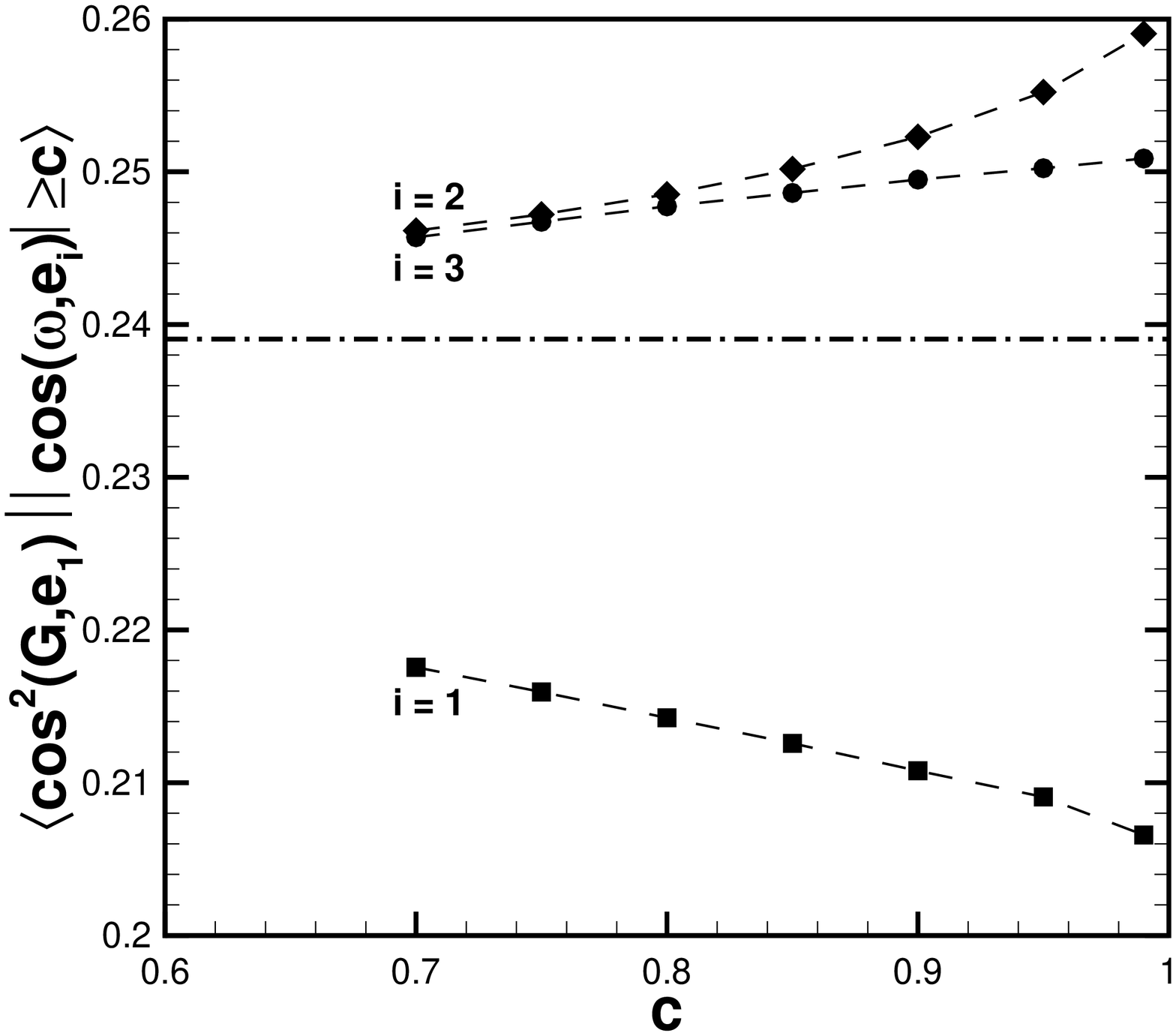}}}%
\subfigure[]{
\resizebox*{7cm}{!}{\includegraphics{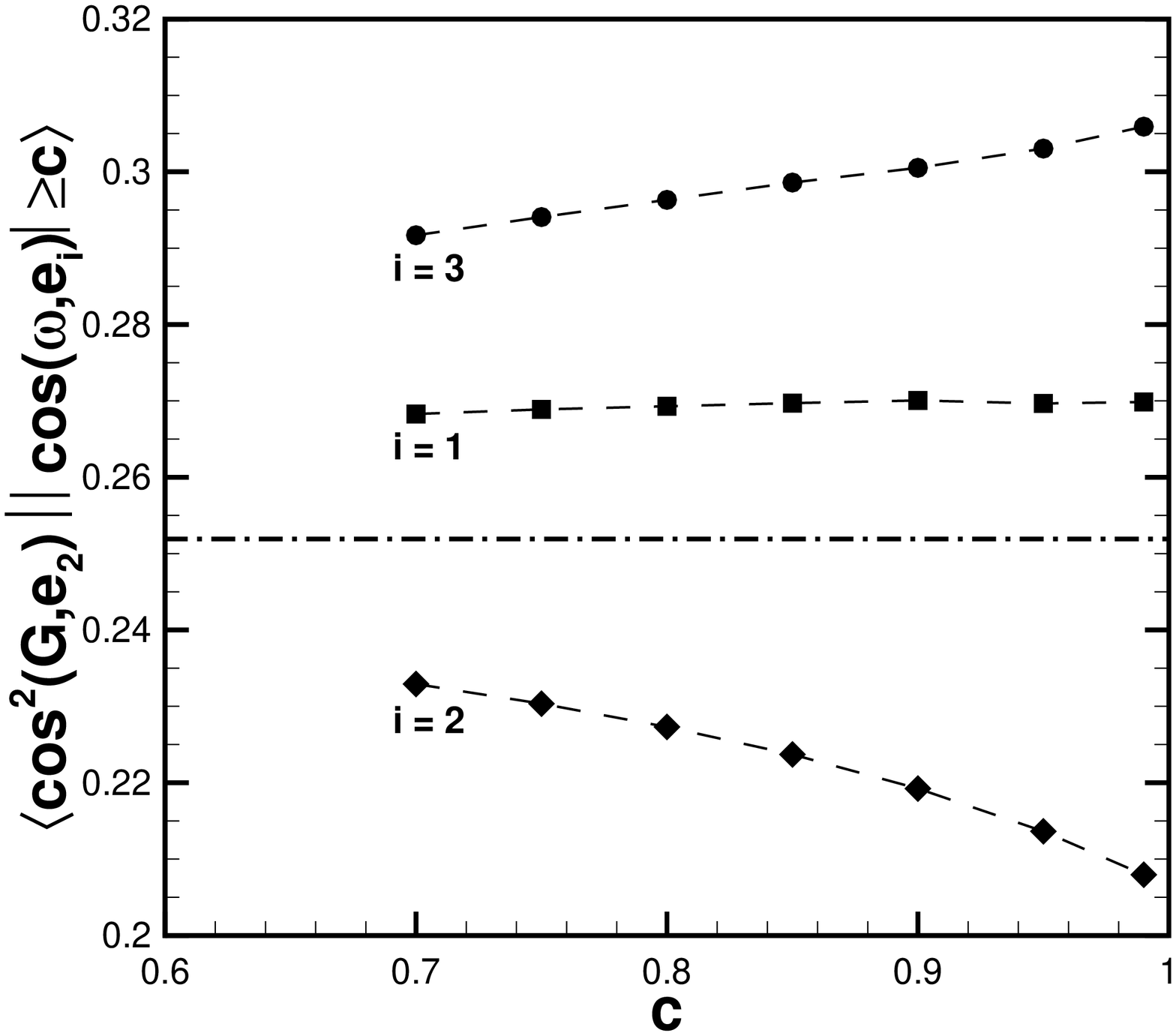}}}%
\end{minipage}
\begin{minipage}{140mm}
\begin{center}
\subfigure[]{
\resizebox*{7cm}{!}{\includegraphics{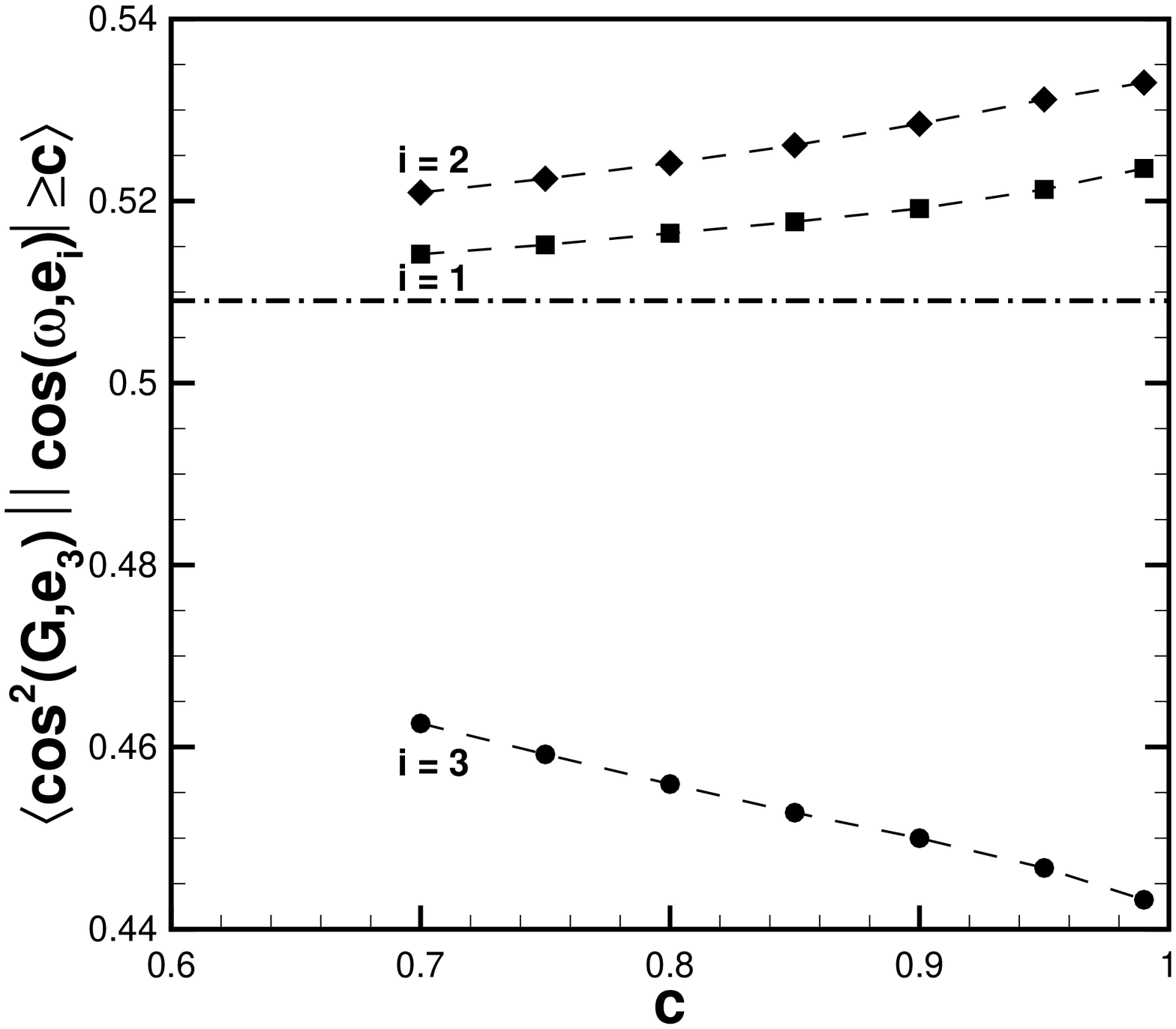}}}%
\end{center}
\caption{Mean square of director cosines of scalar gradient
conditioned on vorticity alignments
with respect to the strain eigenvectors, $ {\bm e}_i $;
angle with respect to
(a) the extensional strain eigenvector,
(b) the intermediate strain eigenvector and
(c) the compressional strain eigenvector;
dash-dotted line: unconditioned average.}%
\label{fig8}
\end{minipage}
\end{center}
\end{figure}

The above picture is explained by 
both 
the respective intensities of the strain components
and the alignments of the scalar gradient with respect
to the strain principal axes 
(Figs. \ref{fig7} and \ref{fig8}).

When $ \bm{\omega} // \bm{e}_3 $ the scalar gradient
alignment with both $ \bm{e}_1 $ and $ \bm{e}_2 $ is rather
good, but the weak values of the intensities of
the extensional and the intermediate strain result in a weak
destruction. However, a small compressional strain intensity
together with a poor alignment of the scalar gradient with
$ \bm{e}_3 $ also bring about a weak production.

\begin{figure}[!h]
\begin{center}
\resizebox*{12cm}{!}{\includegraphics{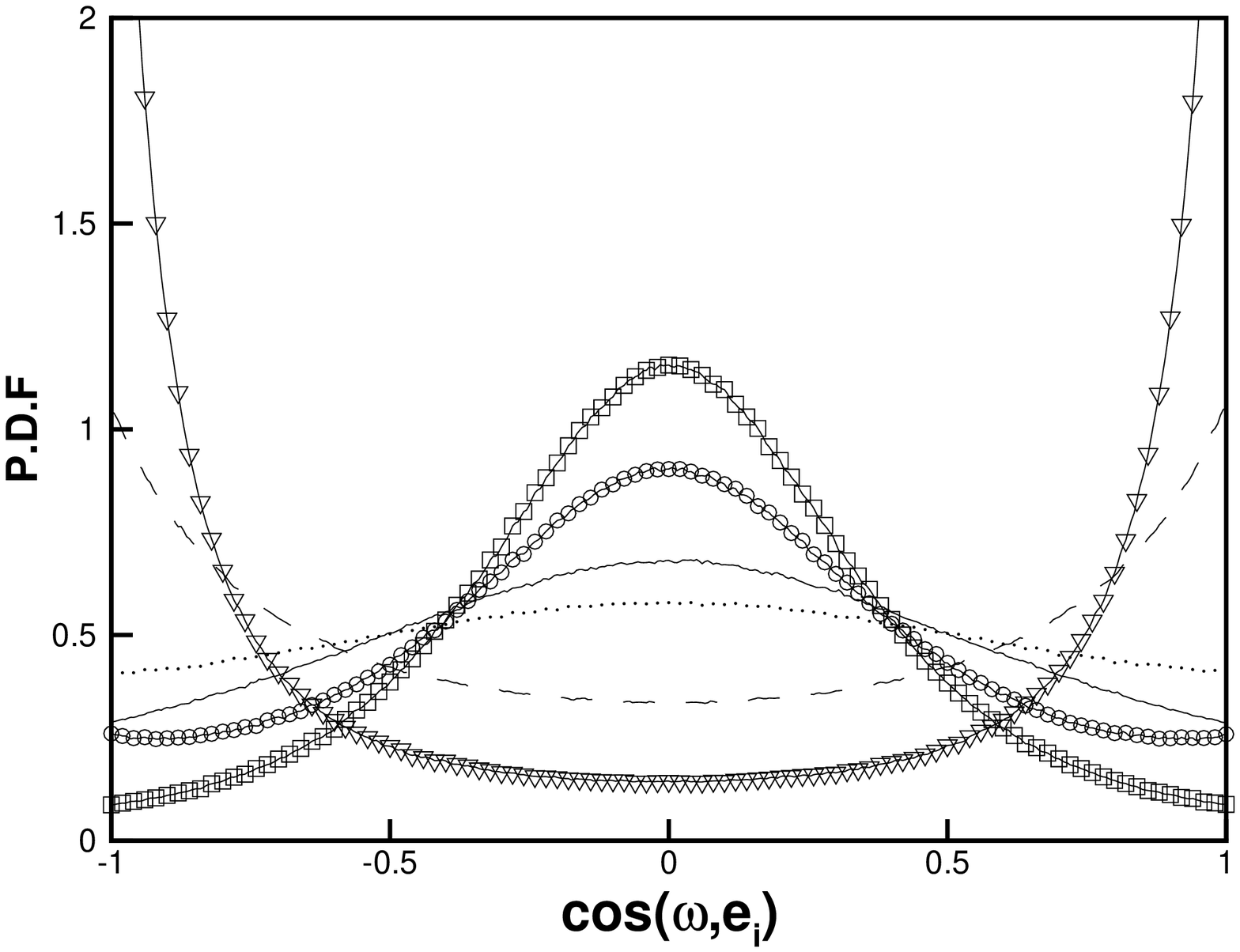}}\\
\caption{P.d.f's of vorticity alignments conditioned on
strain intensity, $ S = {(S_{\alpha \beta} S_{\beta \alpha})}^{1/2} $;
lines are for $ S^2 < \langle S^2 \rangle $;
dotted: $ i = 1 $, dashed: $ i = 2 $, solid: $ i = 3 $;
symbols are for $ S^2 > \langle S^2 \rangle $;
circles: $ i = 1 $, gradients: $ i = 2 $, squares: $ i = 3 $;
the figure compares well with Fig. 9(d) of reference \cite{Lal05}.}
\label{fig9}
\end{center}
\end{figure}

With regard to the difference in scalar gradient production when 
$ \bm{\omega} // \bm{e}_1 $ or $ \bm{\omega} // \bm{e}_2 $:
the extensional strain eigenvalue is the largest and alignment of
the scalar gradient with $ \bm{e}_1 $ is the best for $ \bm{\omega} // \bm{e}_2 $
which explains the largest destruction in this case;
however,
for $ \bm{\omega} // \bm{e}_2 $,
the compressional strain eigenvalue is the largest
and alignment of $ \bm{G} $
with $ \bm{e}_3 $ is
slightly better
which makes 
production larger when $ \bm{\omega} // \bm{e}_2 $
than when $ \bm{\omega} // \bm{e}_1 $; furthermore, because the difference
in the statistics of the intermediate strain eigenvalue between cases
$ \bm{\omega} // \bm{e}_1 $ and $ \bm{\omega} // \bm{e}_2 $ is small
and $ \bm{G} $ aligns better with $ \bm{e}_2 $ for $ \bm{\omega} // \bm{e}_1 $,
destruction by the intermediate strain is larger when $ \bm{\omega} // \bm{e}_1 $.
Better alignment of $ \bm{G} $ with $ \bm{e}_1 $ for
$ \bm{\omega} // \bm{e}_2 $
-- and with $ \bm{e}_2 $
for $ \bm{\omega} // \bm{e}_1 $ --
results from the trend of the scalar gradient and vorticity to
be normal to each other
\cite{Bal03,Gal07}
which is predicted by the model \cite{G09}.
In addition, greater absolute values of the
$ \langle \lambda_i \rangle $'s for  
$ \bm{\omega} // \bm{e}_2 $
are consistent with the fact that moderate and strong production
of strain occurs 
when vorticity 
strongly aligns with
the intermediate strain eigenvector
and 
is rather misaligned with respect to
the extensional 
strain
\cite{T00}.
It has also been shown \cite{Lal05,S91} that
strong alignment of vorticity with 
$ \bm{e}_2 $
is 
correlated with large strain intensity.
The model reproduces the latter property as shown in
Fig. \ref{fig9} which compares rather well with
Fig. 9(d) of reference \cite{Lal05}.

To summarize,
this analysis suggests that the budget of scalar gradient production resulting
in a more intense scalar dissipation for $ \bm{\omega} // \bm{e}_2 $
than for $ \bm{\omega} // \bm{e}_1 $
is explained by the following mechanisms: both extensional strain intensity 
and scalar gradient alignment explain the difference in the destruction
of scalar gradient by the extensional strain, while alignment of the scalar
gradient is the main mechanism resulting in a difference in destruction by
the intermediate strain; and the difference in the production by compressional
strain is to be essentially put down to the compressional strain intensity.

\subsection{Analysis in terms of local flow structure}
\label{sec4.2}

Strain persistence was originally defined in
two-dimensional flows \cite{Lal99,TK94}. It can be extended 
to the three-dimensional case \cite{G09,GG06}
when vorticity is closely aligned with a strain eigenvector
and used
to check whether the flow is locally strain- or rotation-dominated.
The strain persistence parameters are computed for
$ |\cos(\bm{\omega},\bm{e}_i)| \geq 0.99 $ $ (i = 1,2,3) $ and 
are respectively given by:
$ r_1 = - \omega'_1/(\lambda_2 - \lambda_3) $, 
$ r_2 = \omega'_2/(\lambda_1 - \lambda_3) $ 
and
$ r_3 = \omega'_3/(\lambda_1 - \lambda_2) $ 
with
$ \omega'_i = \hat{\omega}_i - \Omega_i $;
the $ \Omega_i $'s are the components of the rotation rate
of strain principal axes computed as:
$ \Omega_1 = -2 \hat{\Pi}_{23}/(\lambda_2 - \lambda_3) $,
$ \Omega_2 = 2 \hat{\Pi}_{13}/(\lambda_1 - \lambda_3) $
and
$ \Omega_3 = -2 \hat{\Pi}_{12}/(\lambda_1 - \lambda_2) $
where the $ \Pi_{ij} $'s are the components of the pressure
Hessian tensor,
$ \Pi_{ij} = (1/\rho) \partial^2 p/\partial x_i \partial x_j $
-- with $ \rho $ and $ p $ standing for density and pressure,
respectively --
modelled as shown in Section \ref{sec2.1}.
Hatted quantities indicate components in the
strain basis.
Prevailing strain is defined by $ r_i^2 < 1 $, while
$ r_i^2 > 1 $ indicates prevailing rotation.
By including the rotation rate of strain principal axes
in addition to vorticity,
strain persistence considers the effective rotation
rate
and was shown to give a better 
estimate of local stirring properties than criteria 
just allowing for vorticity \cite{Lal99}.

The mechanisms of scalar gradient production can be
analysed in terms of prevailing strain {\em vs.} prevailing
rotation from Table \ref{tab1}.
Within the text conditioned mean values such as
$ \langle \bm{G}^2
|\bm{\omega}//\bm{e}_i ; r_i \rangle $
are denoted by $ \langle \bm{G}^2 \rangle_c $.

\begin{table}[!h]
  \tbl{Averaged quantities relevant to scalar gradient production
       conditioned 
       on vorticity alignment
       ($ |\cos(\bm{\omega},\bm{e}_i)| \geq 0.99 $)
       and strain persistence
       parameters, $ r_i $;
       $ r_i^2 < 1 $: prevailing strain;
       $ r_i^2 > 1 $: prevailing rotation;
       bold numbers represent the unconditioned mean values.}
{\begin{tabular}{@{}cccccccc}
\toprule
   strain persistence
        & $ r_1^2 < 1 $
        & $ r_1^2 > 1 $
        & $ r_2^2 < 1 $
        & $ r_2^2 > 1 $
        & $ r_3^2 < 1 $
        & $ r_3^2 > 1 $
        & \\
\colrule
   \% in each $ \bm{\omega}//\bm{e}_i $-sample
     & 38\% & 62\% & 49\% & 51\% & 9.0\% & 91\% \\
\colrule
   $ \langle \bm{G}^2
     |\bm{\omega}//\bm{e}_i ; r_i \rangle $
     & 7.95 & 5.39 & 8.22 & 6.85 & 4.89 & 4.27 & $ \bm{5.15} $ \\
\colrule
   $ \langle \lambda_1
     |\bm{\omega}//\bm{e}_i ; r_i \rangle $
     & 1.87 & 1.55 & 2.58 & 2.52 & 1.94 & 1.67 & $ \bm{1.98} $ \\
   $ \langle \lambda_2
     |\bm{\omega}//\bm{e}_i ; r_i \rangle $
     & 0.979 & 0.424 & 0.822 & 0.523 & 0.00269 & 0.278 & $ \bm{0.534} $ \\
   $ \langle \lambda_3
     |\bm{\omega}//\bm{e}_i ; r_i \rangle $
     & $ -2.86 $ & $ -1.98 $ & $ -3.40 $ & $ -3.04 $ & $ -1.94 $ &  $ -1.95 $ 
     & $ \bm{-2.52} $ \\
\colrule
   $ \langle \cos^2(\bm{G},\bm{e}_1) 
     |\bm{\omega}//\bm{e}_i ; r_i \rangle $
     & 0.186 & 0.220 & 0.236 & 0.281 & 0.225 & 0.253 & $ \bm{0.239} $ \\
   $ \langle \cos^2(\bm{G},\bm{e}_2)
     |\bm{\omega}//\bm{e}_i ; r_i \rangle $
     & 0.229 & 0.296 & 0.201 & 0.215 & 0.295 & 0.307 & $ \bm{0.252} $ \\
   $ \langle \cos^2(\bm{G},\bm{e}_3)
     |\bm{\omega}//\bm{e}_i ; r_i \rangle $
     & 0.585 & 0.485 & 0.563 & 0.505 & 0.480 & 0.440 & $ \bm{0.509} $ \\
\colrule
   $ \langle - \lambda_1 \bm{G}^2 \cos^2(\bm{G},\bm{e}_1)
     |\bm{\omega}//\bm{e}_i ; r_i \rangle $
     & $ -1.22 $ & $ -1.07 $ & $ -4.08 $ & $ -4.70 $ & $ -1.35 $ & $ -1.33 $ 
     & $ \bm{-2.21} $ \\
   $ \langle - \lambda_2 \bm{G}^2 \cos^2(\bm{G},\bm{e}_2)
     |\bm{\omega}//\bm{e}_i ; r_i \rangle $
     & $ -1.06 $ & $ -0.608 $ & $ -0.555 $ & $ -0.337 $ & $ 0.103 $ & $ -0.174 $
     & $ \bm{-0.444} $ \\
   $ \langle - \lambda_3 \bm{G}^2 \cos^2(\bm{G},\bm{e}_3)
     |\bm{\omega}//\bm{e}_i ; r_i \rangle $
     & 19.7 & 8.04 & 22.0 & 14.7 & 7.01 & 5.38 & $ \bm{12.0} $ \\
   net production
     & 17.4 & 6.36 & 17.4 & 9.66 & 5.76 & 3.88 & $ \bm{9.35} $ \\
\botrule
\end{tabular}}
\label{tab1}
\end{table}

In the
$ \bm{\omega}//\bm{e}_2 $-sample 
strain is as frequent as rotation,
while
the
$ \bm{\omega}//\bm{e}_1 $-sample 
is slightly
rotation-dominated
which is consistent with a larger scalar dissipation
for
$ \bm{\omega}//\bm{e}_2 $
than for
$ \bm{\omega}//\bm{e}_1 $.
This result also
suggests that strain is statistically more persistent when vorticity
aligns with the intermediate strain eigenvector and may explain the slightly
better alignment with the compressional strain direction
in this case (Fig. \ref{fig8}). The 
$ \bm{\omega}//\bm{e}_3 $ sample,
by contrast, is found to be strongly rotation-dominated;
in these compressed-vorticity events
small compressional
strain intensity together with a poor alignment of the scalar gradient with
$ \bm{e}_3 $
result in low levels of production
for both prevailing strain and rotation.

As expected,
the largest values of
$ \langle \bm{G}^2 \rangle_c $
as well as the largest production
are found 
for $ \bm{\omega} // \bm{e}_1 $
and
$ \bm{\omega} // \bm{e}_2 $.
In agreement with previous studies \cite{Bal03,Gal07},
production 
mainly results from
prevailing-strain
events.
However, 
$ \langle \bm{G}^2 \rangle_c $
is significant for prevailing rotation as it takes
values greater than the unconditioned average
for both
$ \bm{\omega} // \bm{e}_1 $
and
$ \bm{\omega} // \bm{e}_2 $.
In addition, the difference between
$ \bm{\omega} // \bm{e}_1 $
and
$ \bm{\omega} // \bm{e}_2 $
clearly arises from the rotation-dominated events;
the difference in
$ \langle \bm{G}^2 \rangle_c $
is indeed 3.4\% for
$ r_1^2 < 1 $ 
and
$ r_2^2 < 1 $
-- respectively, 7.95 and 8.22 --, while it reaches
27\% for
$ r_1^2 > 1 $ 
and
$ r_2^2 > 1 $
-- respectively, 5.39 and 6.85.
For $ \bm{\omega} // \bm{e}_3 $, $ \langle \bm{G}^2 \rangle_c $
is smaller than its unconditioned value       
for both prevailing strain and rotation.

The net production confirms the role of rotation
events in the difference between cases
$ \bm{\omega} // \bm{e}_1 $
and
$ \bm{\omega} // \bm{e}_2 $:
the same amount, 17.4, is found for prevailing strain,
while for prevailing rotation the net production
is equal to
6.36 and 9.66, respectively, namely a difference
as large as 52\%.
As
the differences in
total destruction by the extensional
and
the intermediate strains are of the same order for both prevailing
strain and prevailing rotation,
the large difference in net production 
for prevailing rotation
is mainly explained
by
production resulting from
the compressional strain:
12\% for
prevailing strain
-- 19.7 and 22.0 for
$ \bm{\omega} // \bm{e}_1 $
and
$ \bm{\omega} // \bm{e}_2 $, respectively -- and
83\% for prevailing rotation -- 8.04 and 14.7, respectively.
It is also worth noting that production for 
prevailing rotation when
$ \bm{\omega} // \bm{e}_2 $
is not insignificant since both production by the compressional
strain -- 14.7 -- and the net production -- 9.66 -- are greater 
than the corresponding unconditioned values, 12.0 and 9.35,        
respectively.

The mechanisms put forward in Section
\ref{sec4.1} 
to explain the difference in scalar gradient production 
between 
$ \bm{\omega} // \bm{e}_1 $
and
$ \bm{\omega} // \bm{e}_2 $
are emphasized by prevailing rotation.
The mean values of the extensional and compressional
strains, $ \lambda_1 $ and $ \lambda_2 $,
are greater 
when
$ \bm{\omega} // \bm{e}_2 $
than
when
$ \bm{\omega} // \bm{e}_1 $
for both prevailing strain and rotation, but
the difference is larger for prevailing rotation:
for $ \langle \lambda_1 \rangle_c $ the difference
is 38\% for prevailing strain -- respectively, 1.87 and
2.58 -- and 63\% for prevailing rotation -- respectively
1.55 and 2.52;
for 
$ \langle \lambda_3 \rangle_c $ these differences
are respectively 19\% and 54\%.
Although the difference in the alignment of $ \bm{G} $ with respect to
$ \bm{e}_1 $ is almost the same for prevailing strain and prevailing
rotation 
-- 27\% and 28\% --,
the difference in the alignment with $ \bm{e}_2 $ is larger
for prevailing rotation
-- 14\% and 38\%.

\section{Conclusion}

\label{sec5}
Scalar dissipation has been analysed through scalar gradient production
using a stochastic Lagrangian model for the velocity and the scalar
gradients which reproduces the essential dynamic and kinematic properties
of isotropic turbulence. The study was specifically focused on the connection
between vorticity geometry and scalar dissipation, and thus small-scale
mixing.

The model results show that scalar dissipation is mainly found when vorticity
is stretched. More precisely, for vorticity aligning with the extensional
strain, scalar dissipation is close to its unconditioned mean value, while
the most intense scalar dissipation -- and therefore the most efficient
small-scale mixing -- occurs
for vorticity aligning with the intermediate strain.
Scalar dissipation is significantly lower when vorticity is compressed.

The difference in scalar dissipation when vorticity aligns with either the
extensional -- $ \bm{\omega} // \bm{e}_1 $ -- 
or the intermediate strain -- $ \bm{\omega} // \bm{e}_2 $ --
is to be put down 
to the interplay of mechanisms involving the strain intensities and the
alignment of the scalar gradient with respect to the strain principal
axes. In brief, for $ \bm{\omega} // \bm{e}_2 $
both a larger extensional strain intensity and a better alignment
of the the scalar gradient with the extensional strain result in
a larger destruction of the scalar gradient norm than when 
$ \bm{\omega} // \bm{e}_1 $;
however, this effect is exceeded
by production caused by compression, essentially through a larger
compressional strain intensity; in addition, when $ \bm{\omega} // \bm{e}_2 $ 
it is mainly the misalignment
of the scalar gradient with $ \bm{e}_2 $ that causes a lesser destruction by the intermediate
strain.
For vorticity aligning with the compressional strain direction
-- $ \bm{\omega} // \bm{e}_3 $ --,
weak production of scalar gradient results from both
small intensity of the compressional strain and misalignment 
between
the scalar gradient 
and
$ \bm{e}_3 $.

Finally, 
the
latter
mechanisms,
and especially those
explaining the difference in scalar gradient
production between
$ \bm{\omega} // \bm{e}_1 $
and
$ \bm{\omega} // \bm{e}_2 $,
are retrieved in the analysis in terms of local flow structure.
Although scalar gradient production mostly stems from
prevailing-strain events,
it appears that this difference is the largest
in rotation-dominated events, especially with regard to
the extensional and compressional strain intensities.

\end{document}